\documentclass[oldversion]{aa}
\usepackage{epsfig,graphicx}

\usepackage{natbib}
\usepackage[normalem]{ulem}

\begin{document}

\title{Drastic changes in the molecular absorption at redshift $z=0.89$ toward the quasar PKS~1830$-$211}

\author{Muller S. \inst{1} \and Gu\'elin M. \inst{2,3}}

\offprints{muller@asiaa.sinica.edu.tw}
\institute{
Academia Sinica Institute of Astronomy and Astrophysics (ASIAA),
P.O. Box 23--141, Taipei, 106, Taiwan
\and
Institut de Radio Astronomie Millim\'etrique (IRAM), 300 rue de la piscine, F--38406 St Martin d'H\`eres,
\and
Ecole Normale Sup\'erieure/LERMA, 24 rue Lohmond, F--75005 Paris, France}

\date {Received 16 June 2008 / Accepted 24 September 2008}
%\thesaurus{}

\titlerunning{Changes in the $z=0.89$ molecular absorption toward PKS~1830$-$211}
\authorrunning{Muller \& Gu\'elin 2008}

\abstract{ A 12 year-long monitoring of the absorption caused by a $z=0.89$ spiral galaxy
on the line of sight to the radio-loud gravitationally lensed quasar PKS~1830$-$211 reveals
spectacular changes in the HCO$^+$ and HCN (2-1) line profiles. The depth of the absorption
toward the quasar NE image increased by a factor of $\sim 3$ in 1998-1999 and subsequently
decreased by a factor $\geq 6$ between 2003 and 2006. These changes were echoed by similar
variations in the absorption line wings toward the SW image. Most likely, these variations
result from a motion of the quasar images with respect to the foreground galaxy,
which could be due to a sporadic ejection of bright plasmons by the background quasar.
VLBA observations have shown that the separation between the NE and SW images changed in 1997 by
as much as 0.2 mas within a few months. Assuming that motions of similar amplitude
occurred in 1999 and 2003, we argue that the clouds responsible for the NE absorption and
the broad wings of the SW absorption should be sparse and have characteristic sizes of
$0.5-1$ pc.

\keywords{Quasars: individual: PKS~1830$-$211 -- Quasars: absorption lines -- Galaxies: ISM -- ISM: molecules}}

\maketitle

\section{Introduction}

Atomic and molecular line absorption measurements against bright sources are a powerful
tool for studying the chemical composition and structure of the interstellar gas. In the cases
where very small structures are present and where, due to proper motions, the line of sight
scans across these structures, the absorption profile may vary with time. Such time
variations allow the cloud properties to be studied on scales hardly accessible to other
techniques.

Time variations of Galactic HI 21-cm line absorptions have been reported on the line of sight
to pulsars by \cite{fra94}, but were not reproduced by later studies (e.g., \citealt{sta03}).
\cite{bro05} find temporal variations of the HI line absorption on the line of sight of the
quasar 3C~138 over a 7-yr time span, although only out of three epochs of data.
Time variations have also been observed for optical atomic lines toward bright stars [see a
review in \citealt{cra03}]. Typically, these variations have amplitudes of 10\% or less and time
scales of several months or years. They are interpreted as the signature of structures as small as
$10-100$ astronomical units (AU) in the Galactic interstellar medium (ISM). Evidence of such
small structures comes from VLBI absorption measurements against closely packed continuum sources
(\citealt{die76,dia89,fai01}). Most likely, they are clumps or filaments expelled by supernova
explosions or stellar winds.

As for molecular gas, variations have been detected in centimeter-wave absorption spectra of H$_2$CO
(\citealt{mar93}) and OH (\citealt{moo95}), in millimeter-wave spectra of HCO$^+$ (\citealt{lis00})
and in optical spectra of CH, CH$^+$ and CN (\citealt{pan01,rol03}). Here also, the absorption occurs
in Galactic diffuse clouds seen against bright stars or quasars. The variations are observed over
periods $>1$ yr and their amplitude is seldom above 10\%. They arise from $10-100$ AU size structures
of the Galactic ISM similar to those observed in the atomic lines.

There are less data concerning absorption variations in extragalactic clouds. Time variability of
HI line absorption has been reported at $z=0.52$ (\citealt{wol82}), $z=0.31$ (\citealt{kan01}) and
$z=0.67$ (\citealt{kan08}) in front of radio quasars. The short time scales, on the order of a few
days for the first two studies, may be explained by interstellar scintillation coupled with parsec
or sub-parsec scale structures in the absorbing galaxies.

\cite{wik97} also report time variations in the CO absorption profile observed against the quasar
PKS~1413+135 at a redshift $z=0.25$. The absorption occurs in the host galaxy, which is
seen almost edge-on, and the variations primarily affect the nearly saturated absorption
peak, the intensity of which was found to decrease over two years. \cite{wik97} interpret
these variations by structural changes in the background continuum source, resulting in
a changing line of sight through the absorbing gas. The latter consists of small dense clumps
embedded in a mostly diffuse medium.

In this article, we report the detection of large variations in the HCO$^+$ and HCN
absorption profiles observed at $z=0.89$ on the line of sight to the quasar PKS~1830$-$211.
PKS~1830$-$211 is a radio-loud quasar at a redshift of $z=2.5$ (\citealt{lid99}), whose
line of sight is intercepted by a spiral galaxy (\citealt{win02}) at a redshift $z=0.88582$
(\citealt{wik96}). The galaxy gives rise to absorption in many molecular lines (see, e.g.,
\cite{wik98,mul06} --hereafter MGD-- and references therein) and acts as a gravitational
lens. At millimeter wavelengths, the image of the quasar is split into two compact
continuum sources (denoted NE and SW images), separated by $1\arcsec$, or $\simeq 8$ kpc,
and they appear on opposite sides of the galaxy bulge. At $\lambda=3$ mm, the NE image
contributes about 2/3 of the total continuum flux and the SW image the remaining
1/3, a ratio that fluctuates in relation to background source flux variations
(\citealt{van95,wik99}) coupled with a time delay of about 25 days (\citealt{lov98,wik99}).

This peculiar configuration makes it possible to explore along two pencil beams the
chemical composition and structure of the gas. Absorption is most prominent in the $J=2-1$
lines of HCO$^+$ and HCN, both of which show two distinct absorption components: {\it i)} a
broad component (FWZP $\Delta V = 100$ km~s$^{-1}$) around $V = 0$ km~s$^{-1}$ {\footnote
{We adopt a heliocentric reference frame.}}, associated with the SW image, which is
obviously saturated and presumably consists of several distinct clouds, and {\it ii)} a
narrow component ($\Delta V \sim 15$ km~s$^{-1}$), 146 km~s$^{-1}$ lower in velocity and
associated with the NE image. Both components were found to vary in intensity and shape in
the course of a 12 year-long monitoring of the quasar. The amplitude of the variations, the
duration of the monitoring period and the redshift of the absorbing system offer a rare
opportunity to investigate the gas properties in the arms of a relatively young spiral
galaxy.

\begin{figure*}[t]
\includegraphics[width=18cm]{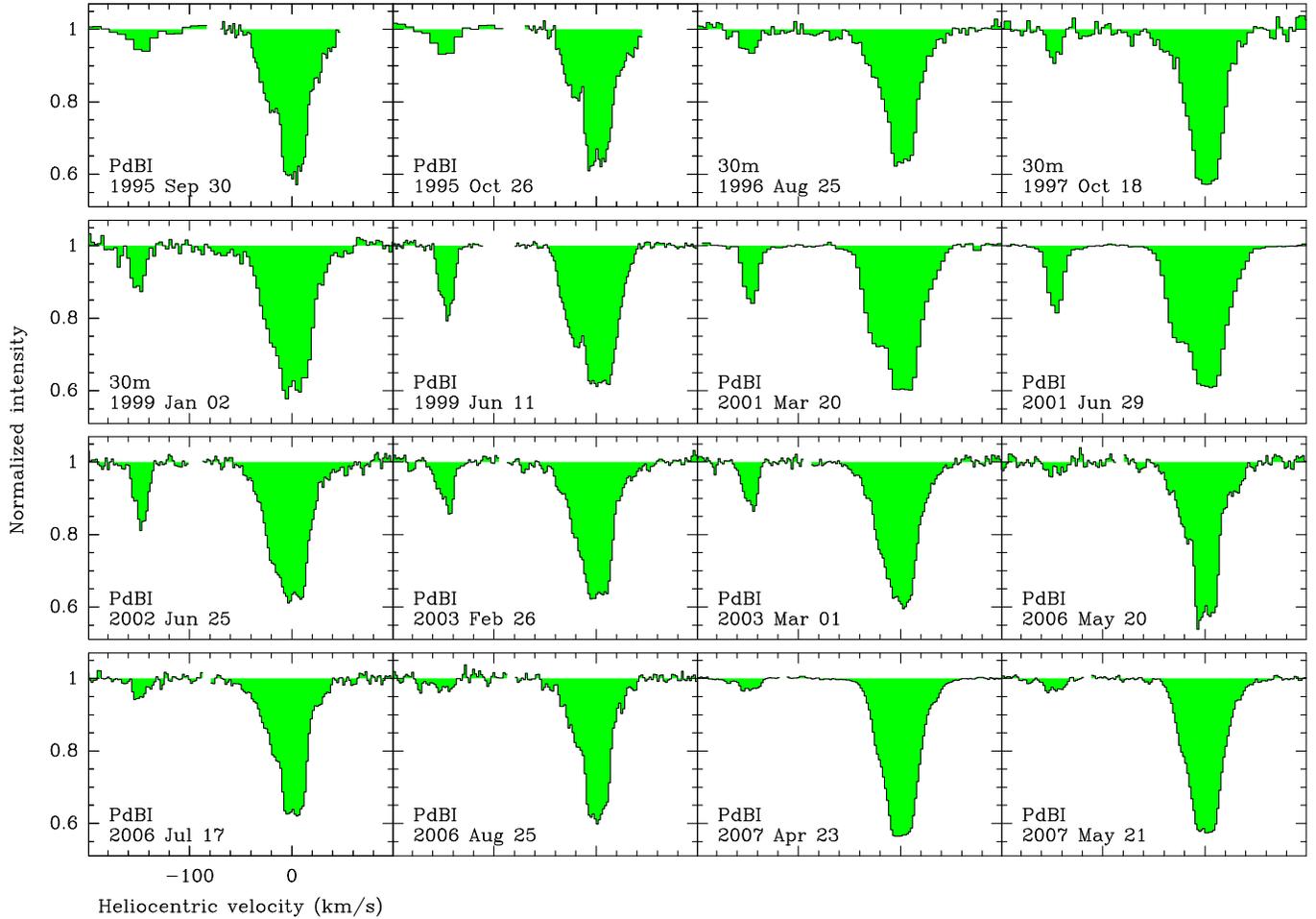}
\caption{Spectra of the HCO$^+$ absorption toward PKS~1830$-$211, obtained either with the PdBI or the 30m,
at different epochs between 1995 and 2007. The intensity is normalized with respect to the total (NE+SW) continuum flux.
The velocity resolution is 2 km~s$^{-1}$, except for the spectral window on the $-146$ km~s$^{-1}$
component in 1995, where it is 8 km~s$^{-1}$, for 2001 spectra, where it is 4 km~s$^{-1}$, and for 30m spectra,
where it is 3 km~s$^{-1}$.}
%The spectral window on the 0 km~s$^{-1}$ component
%in the HCO$^+$ 1996 spectrum was affected by a bandpass problem. The HCO$^+$ and HCN spectra in 2007
%were observed simultaneously.}
\label{spec-hco}
\end{figure*}

\begin{figure*}[t]
\includegraphics[width=18cm]{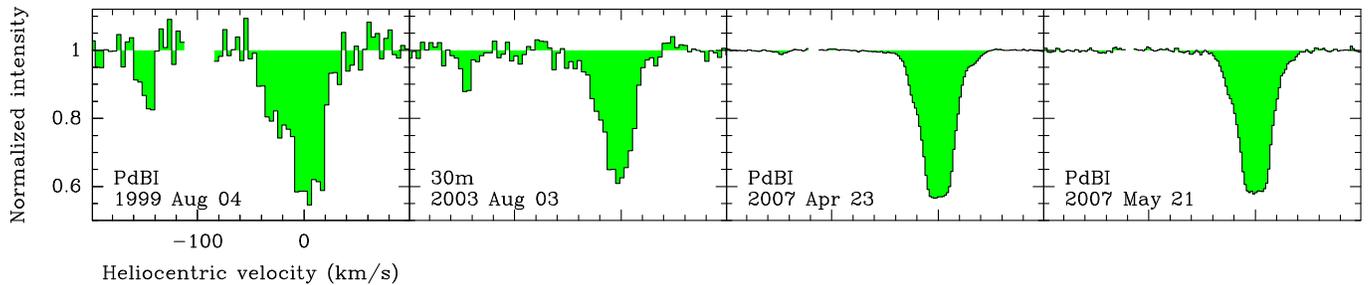}
\caption{Spectra of the HCN (2-1) absorption toward PKS~1830$-$211 at different epochs between 1999 and 2007.
The velocity resolution is 4 km~s$^{-1}$ for 1999 and 2003 spectra and 2 km~s$^{-1}$ for 2007 spectra.}
%The 2003 spectrum was observed with the IRAM 30m.}
\label{spec-hcn}
\end{figure*}

\begin{figure*}[t]
\includegraphics[width=18cm]{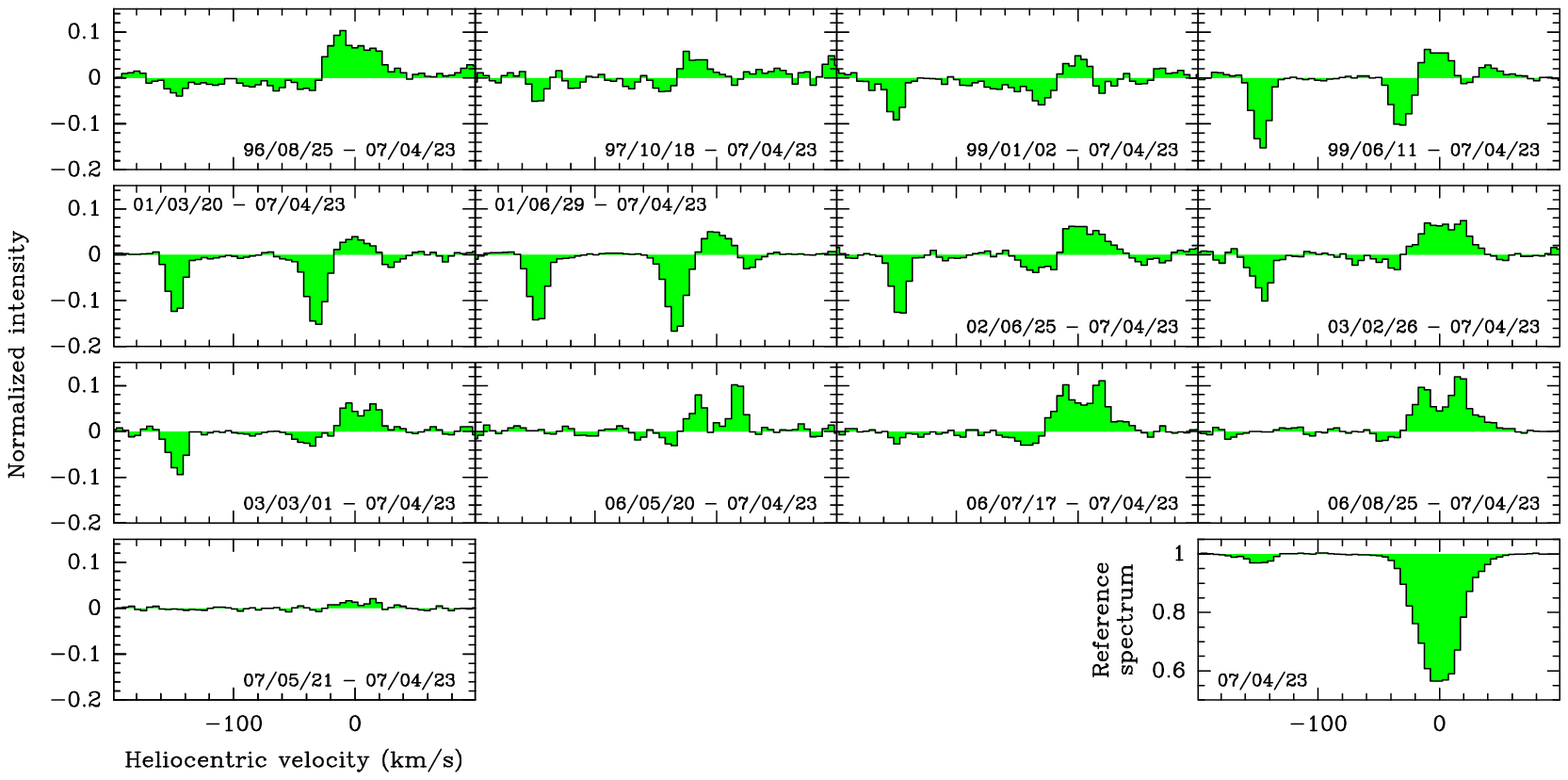}
\caption{Difference spectra for the HCO$^+$ (2-1) line.
The reference spectrum is that of 2007 Apr 23 ({\em bottom right}). The date format is yy/mm/dd.
Spectra were resampled and smoothed to a velocity resolution of 5 km~s$^{-1}$.}
\label{diffspectra}
\end{figure*}

\section{Observations}

The observations discussed in this paper concern the HCO$^+$ and HCN ($J=2-1$) lines,
redshifted by 0.88582 to $\sim 94$ GHz, whose absorption profiles were monitored over the
period September 1995 -- May 2007. They were carried out with the IRAM Plateau de Bure
interferometer (PdBI) or the IRAM 30m telescope.

The PdBI data were mostly taken with the array in a compact configuration, often in average
weather conditions, in the frame of an observing program whose goal was to measure isotopic
abundance ratios in the lensing galaxy (MGD). The strong flux of PKS~1830$-$211 ($\simeq 2$
Jy at $\lambda = 3$ mm) allowed us to self-calibrate the spectra with respect to the
continuum emission, so that accurate absorption profiles, normalized to the continuum,
could be derived even under marginal atmospheric phase stability. The calibration accuracy
on the normalized PdBI spectra is better than a percent of the continuum level. The HCO$^+$
(2-1) spectra obtained with the PdBI in the period 1995 -- 2007, normalized to the total
(NE+SW) continuum flux, are displayed on Fig.\ref{spec-hco}. The integration time for each
individual spectrum ranged from 0.5 to 3 hours and the signal-to-noise ratio is generally
excellent.

Additional HCO$^+$ (2-1) spectra were observed between 1996 and 1999 by T. Wiklind and F.
Combes with the IRAM 30-m telescope. They were part of a program aiming to measure the
Hubble constant H$_0$ using the variability of the quasar and the geometrical time delay
between the two gravitationally lensed images. As it is difficult to concisely show all
($\sim 60$) the HCO$^+$ (2-1) 30-m telescope spectra, we have shown in Fig.\ref{spec-hco}
only 3 spectra obtained at critical periods: Aug 1996, Oct 1997 and Jan 1999. The rest of
the 30-m HCO$^+$ data are displayed in the form of flux and the integrated line opacity
data points (see below). A list of all the observations is given in Table~1.

Finally, we collected 4 HCN (2-1) spectra, 3 observed with the PdBI and 1 with the 30-m
telescope. They are shown on Fig.\ref{spec-hcn}. We note that, since 2007, the HCN spectra
were observed simultaneously with the HCO$^+$, thanks to the new broad-band
dual-polarization receivers installed at the PdBI.

Difference spectra of 13 of the HCO$^+$ (2-1) spectra, resampled and smoothed to a velocity
resolution of 5 km~s$^{-1}$, are shown on Fig.\ref{diffspectra}. They clearly illustrate the
drastic changes that affected the absorption profile between 1998 and 2004.

In total, the available data densely cover a 12 year-long period, except for breaks in 1998
and between 2003 and 2006. At each daily session, the flux of PKS~1830$-$211 was derived
from the observation of a strong reference source (3C~273, 3C~345, 1749+096, or MWC~349),
whose flux is regularly monitored with IRAM instruments. These flux measurements are
reported in Fig.\ref{flux}a. We estimate that the continuum flux of
PKS~1830$-$211 should be accurate to 15\% (r.m.s.).

Whereas the knowledge of the NE/SW flux ratio is crucial for the derivation of line opacities,
the low angular resolution of all but the PdBI 2003 observations made it generally impossible to resolve
the NE from the SW image. Nevertheless, the NE/SW continuum flux ratio could accurately
be derived from the depth of the HCO$^+$ and/or HCN (2-1) absorption trough around 0 km~s$^{-1}$, where the
profiles are heavily saturated (see MGD). Since the absorbing gas is very cold with respect to the hot
non-thermal background source, and since the NE source is free of absorption at this velocity, the signal
detected at 0 km~s$^{-1}$ simply reflects the flux of the NE continuum source. We report the evolution
of the flux ratio NE/SW in Fig.\ref{flux}b.

\begin{table}[h] 
\caption{Observations of the HCO$^+$ (2-1) absorption line.} 
\label{tab} 
\begin{center} \begin{tabular}{ccccccc} 
\hline 
Date & Julian Day & Telescope & Total flux $\diamond$ & NE/SW  & $\int \tau$dV (NE) $\dagger$ & $\int \tau$dV (SW blue) $\dagger$ \\ 
     &            &           & (Jy)       & flux ratio $\diamond$      &  (km~s$^{-1}$) & (km~s$^{-1}$) \\ 
\hline 
1995 Sep 30 & 2449991 & PdBI & -- & 1.51  (0.20) & 2.28  (--) & 14.40  (--) \\ 
1995 Oct 28 & 2450017 & PdBI & -- & 1.78  (0.20) & 2.51  (--) & 13.32  (--) \\ 
1996 May 01 & 2450205 & 30m & 1.2 & 1.87  (0.19) & 0.62  (0.78) & 8.17  (1.64) \\ 
1996 May 25 & 2450229 & 30m & 1.1 & 1.84  (0.10) & 1.93  (0.41) & 10.82  (0.84) \\ 
1996 Jun 06 & 2450241 & 30m & 0.8 & 3.13  (0.26) & 1.23  (0.43) & 15.29  (1.53) \\ 
1996 Jun 13 & 2450248 & 30m & 1.0 & 1.89  (0.15) & 2.43  (0.60) & 13.42  (1.27) \\ 
1996 Jun 23 & 2450258 & 30m & 1.0 & 1.79  (0.06) & 2.85  (0.24) & 11.09  (0.47) \\ 
1996 Jun 29 & 2450264 & 30m & 1.0 & 1.65  (0.09) & 2.28  (0.46) & 15.20  (0.84) \\ 
1996 Jul 06 & 2450271 & 30m & 1.1 & 2.99  (0.29) & 1.27  (0.51) & 5.66  (1.78) \\ 
1996 Jul 14 & 2450279 & 30m & 0.9 & 1.52  (0.10) & 2.43  (0.56) & 15.79  (0.94) \\ 
1996 Jul 19 & 2450284 & 30m & 0.9 & 1.50  (0.10) & 2.56  (0.54) & 10.65  (0.90) \\ 
1996 Jul 23 & 2450288 & 30m & 1.0 & 1.71  (0.12) & 0.69  (0.57) & 7.02  (1.09) \\ 
1996 Aug 04 & 2450300 & 30m & 1.0 & 1.63  (0.07) & 2.99  (0.35) & 14.02  (0.64) \\ 
1996 Aug 10 & 2450306 & 30m & 1.2 & 2.18  (0.12) & 1.37  (0.36) & 7.26  (0.88) \\ 
1996 Aug 19 & 2450315 & 30m & 1.1 & 1.73  (0.07) & 1.66  (0.31) & 8.28  (0.60) \\ 
1996 Aug 25 & 2450321 & 30m & 1.0 & 1.75  (0.04) & 1.99  (0.16) & 10.89  (0.30) \\ 
1996 Sep 01 & 2450328 & 30m & 1.1 & 1.38  (0.06) & 1.33  (0.39) & 10.12  (0.59) \\ 
1996 Sep 16 & 2450343 & 30m & 0.8 & 1.08  (0.10) & 2.23  (0.93) & 11.42  (1.12) \\ 
1996 Sep 20 & 2450347 & 30m & 0.8 & 1.9  (0.14) & 3.44  (0.54) & 16.10  (1.13) \\ 
1996 Sep 24 & 2450351 & 30m & 1.0 & 2.19  (0.10) & 2.30  (0.32) & 10.62  (0.78) \\ 
1996 Oct 28 & 2450385 & 30m & 0.9 & 1.37  (0.06) & 2.46  (0.38) & 9.47  (0.58) \\ 
1996 Nov 07 & 2450395 & 30m & 0.9 & 1.55  (0.08) & 2.13  (0.46) & 10.94  (0.79) \\ 
1996 Nov 22 & 2450410 & 30m & 0.9 & 1.77  (0.11) & 3.59  (0.46) & 12.51  (0.90) \\ 
1996 Nov 29 & 2450417 & 30m & 0.9 & 1.64  (0.09) & 1.49  (0.44) & 8.94  (0.79) \\ 
1997 Apr 04 & 2450543 & 30m & 1.0 & 1.83  (0.09) & 0.64  (0.38) & 10.74  (0.78) \\ 
1997 Apr 05 & 2450544 & 30m & 0.9 & 1.7  (0.08) & 2.39  (0.35) & 11.34  (0.67) \\ 
1997 May 18 & 2450587 & 30m & 1.2 & 1.65  (0.05) & 1.23  (0.25) & 9.32  (0.45) \\ 
1997 Jun 13 & 2450613 & 30m & 1.2 & 1.55  (0.06) & 1.20  (0.33) & 10.64  (0.58) \\ 
1997 Jun 28 & 2450628 & 30m & 1.3 & 1.64  (0.07) & 1.63  (0.32) & 10.25  (0.59) \\ 
1997 Jul 10 & 2450640 & 30m & 1.2 & 1.48  (0.04) & 1.86  (0.25) & 9.45  (0.41) \\ 
1997 Jul 13 & 2450643 & 30m & 0.8 & 0.98  (0.09) & 2.62  (1.06) & 8.85  (1.15) \\ 
1997 Aug 09 & 2450670 & 30m & 1.2 & 1.85  (0.12) & 1.41  (0.49) & 13.34  (1.01) \\ 
1997 Aug 12 & 2450673 & 30m & 1.2 & 1.56  (0.07) & 2.09  (0.38) & 10.29  (0.66) \\ 
1997 Aug 13 & 2450674 & 30m & 1.1 & 1.79  (0.19) & -0.06  (0.82) & 10.49  (1.63) \\ 
1997 Aug 15 & 2450676 & 30m & 1.1 & 1.54  (0.10) & 1.28  (0.52) & 11.27  (0.88) \\ 
1997 Aug 16 & 2450677 & 30m & 1.1 & 1.49  (0.20) & 0.95  (1.15) & 5.49  (1.94) \\ 
1997 Aug 19 & 2450680 & 30m & 1.1 & 1.8  (0.13) & 1.91  (0.55) & 7.14  (1.09) \\ 
1997 Aug 22 & 2450683 & 30m & 1.2 & 1.53  (0.09) & 1.54  (0.48) & 11.31  (0.81) \\ 
1997 Aug 26 & 2450687 & 30m & 1.2 & 1.85  (0.10) & 2.06  (0.41) & 9.30  (0.85) \\ 
1997 Aug 30 & 2450691 & 30m & 1.5 & 2.05  (0.11) & 1.76  (0.37) & 10.48  (0.83) \\ 
1997 Sep 03 & 2450695 & 30m & 1.2 & 2.07  (0.08) & 2.00  (0.27) & 8.79  (0.62) \\ 
1997 Sep 06 & 2450698 & 30m & 1.4 & 1.85  (0.10) & 0.91  (0.39) & 7.02  (0.80) \\ 
1997 Sep 10 & 2450702 & 30m & 1.4 & 1.75  (0.07) & 2.00  (0.29) & 8.82  (0.57) \\ 
1997 Sep 16 & 2450708 & 30m & 1.4 & 1.67  (0.10) & 3.14  (0.48) & 12.44  (0.90) \\ 
1997 Sep 26 & 2450718 & 30m & 1.1 & 1.31  (0.07) & 2.33  (0.52) & 10.81  (0.76) \\ 
1997 Oct 03 & 2450725 & 30m & 1.4 & 1.48  (0.07) & 1.62  (0.42) & 12.15  (0.68) \\ 
1997 Oct 18 & 2450740 & 30m & 1.4 & 1.37  (0.05) & 2.25  (0.31) & 10.28  (0.48) \\ 
1998 Dec 25 & 2451173 & 30m & 2.2 & 1.50  (0.07) & 3.53  (0.38) & 16.59  (0.62) \\ 
1999 Jan 02 & 2451181 & 30m & 1.6 & 1.55  (0.04) & 3.35  (0.23) & 17.63  (0.40) \\ 
1999 Jan 11 & 2451190 & 30m & 1.9 & 1.61  (0.05) & 2.97  (0.25) & 15.09  (0.45) \\ 
1999 Jan 27 & 2451206 & 30m & 2.2 & 1.51  (0.06) & 2.48  (0.31) & 14.36  (0.53) \\ 
1999 Jan 30 & 2451209 & 30m & 2.2 & 1.51  (0.06) & 3.17  (0.34) & 14.73  (0.56) \\ 
1999 Apr 23 & 2451292 & 30m & 2.0 & 1.43  (0.04) & 6.01  (0.23) & 20.75  (0.37) \\ 
1999 May 05 & 2451304 & 30m & 2.1 & 1.98  (0.09) & 4.27  (0.34) & 17.91  (0.75) \\ 
1999 May 12 & 2451311 & 30m & 2.0 & 1.63  (0.04) & 5.37  (0.20) & 22.22  (0.36) \\ 
1999 Jun 11 & 2451341 & PdBI & 2.3 & 1.64  (0.02) & 5.6  (0.08) & 22.77  (0.14) \\ 
1999 Jun 13 & 2451343 & 30m & 2.0 & 1.64  (0.05) & 5.64  (0.24) & 22.36  (0.43) \\ 
1999 Jun 18 & 2451348 & 30m & 2.1 & 1.53  (0.04) & 4.70  (0.23) & 21.34  (0.39) \\ 
1999 Jul 09 & 2451369 & 30m & 1.9 & 1.39  (0.04) & 6.07  (0.23) & 22.31  (0.35) \\ 
1999 Jul 23 & 2451383 & 30m & 1.9 & 1.48  (0.04) & 5.81  (0.23) & 25.36  (0.38) \\ 
1999 Jul 30 & 2451390 & 30m & 1.8 & 1.43  (0.04) & 5.70  (0.27) & 24.37  (0.43) \\ 
2001 Mar 20 & 2451989 & PdBI & 2.4 & 1.52  (0.02) & 4.83  (0.10) & 22.76  (0.17) \\ 
2001 Jun 29 & 2452090 & PdBI & 2.9 & 1.58  (0.01) & 5.76  (0.07) & 24.81  (0.12) \\ 
2002 Jun 25 & 2452451 & PdBI & 2.4 & 1.68  (0.04) & 4.75  (0.17) & 20.78  (0.32) \\ 
2003 Feb 26 & 2452697 & PdBI & 1.7 & 1.72  (0.03) & 3.78  (0.14) & 15.10  (0.26) \\ 
2003 Mar 01 & 2452700 & PdBI & 1.8 & 1.60  (0.03) & 3.59  (0.14) & 16.93  (0.25) \\ 
2006 May 20 & 2453876 & PdBI & 2.4 & 1.37  (0.03) & 1.11  (0.19) & 9.68  (0.27) \\ 
2006 Jul 17 & 2453934 & PdBI & 1.7 & 1.71  (0.03) & 1.49  (0.13) & 11.81  (0.23) \\ 
2006 Aug 25 & 2453973 & PdBI & 1.8 & 1.67  (0.03) & 1.00  (0.17) & 9.87  (0.30) \\ 
2007 Apr 23 & 2454214 & PdBI & 2.8 & 1.32  (0.005) & 1.19  (0.04) & 10.51  (0.05) \\ 
2007 May 21 & 2454242 & PdBI & 2.5 & 1.39  (0.01) & 1.26  (0.07) & 10.59  (0.11) \\ 
\hline 
\end{tabular} \end{center}
\mbox{\,} \vskip -.5cm
$\diamond$ see caption of Fig.\ref{flux}; $\dagger$ see caption of Fig.\ref{opa}.
\end{table}

\section{Results}

The most striking result in Fig.\ref{spec-hco} is the change in intensity of the $-146$
km~s$^{-1}$ absorption feature that occurred around 1998 and 2004. The depth of this
feature, which in 1995-1996 amounted to $\simeq 7\%$ of the total continuum (and 11\% of
the NE continuum source), increased by 1999 to 20\% (32\% of the NE continuum source) and
dropped almost to zero ($\leq 5$\%) sometime between 2003 and 2006 -- a drastic change
observed in both HCO$^+$ and HCN (2-1) (see Fig.\ref{spec-hcn}).

\begin{figure}[h]
\includegraphics[width=8cm]{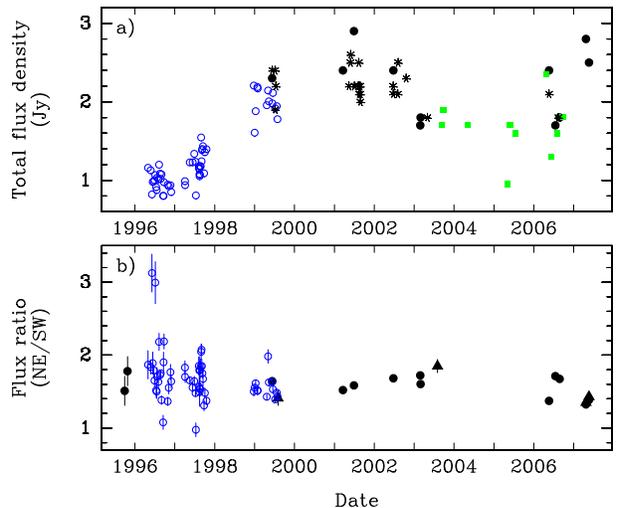}
\caption{{\bf a)} Evolution of the total continuum flux density (NE+SW) of PKS~1830$-$211 at frequencies $\sim 94$ GHz.
Open circles: 30-m measurements at the frequency of the redshifted HCO$^+$ (2-1) line;
filled circles: same for PdBI measurements;
asterisks: PdBI measurements at frequencies between 91 and 104 GHz; % and with phase decorrelation $< 50\deg$;
triangles: PdBI measurements for HCN (2-1);
squares: flux measurements with the Australian Telescope Compact Array (from the webpage:
http://www.narrabri.atnf.csiro.au/calibrators/) %data/1830-210/ta\_w.txt}
at various frequencies between 89 and 96 GHz.
{\bf b)} Evolution of the flux ratio NE/SW, as derived from the average depth of the absorption between
$-7$ and +12 km~s$^{-1}$, where the signal represents the continuum flux of the NE source (see text).}
\label{flux}
\end{figure}

\begin{figure}[h]
\includegraphics[width=8cm]{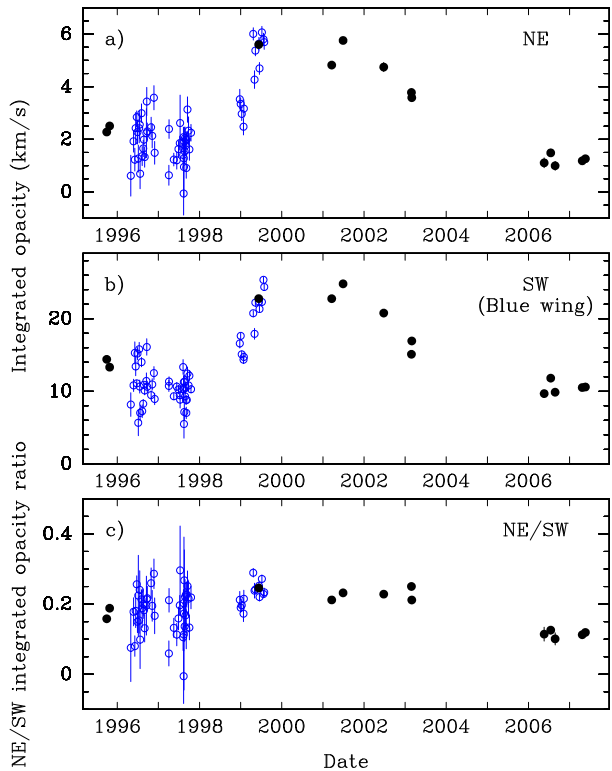}
\caption{Variations of {\bf a)} the integrated opacity of the NE absorption component
(integrated between $-160$ and $-130$ km~s$^{-1}$), {\bf b)} the integrated opacity of the
blue wing of the SW component (integrated between $-50$ and $-15$ km~s$^{-1}$), and
{\bf c)} the ratio of the two previous quantities.
Open circles correspond to 30m data and filled circles to PdBI observations. See Table 1.}
\label{opa}
\end{figure}

The changes in the NE image absorption are echoed in the SW image
by parallel changes of the blue absorption wing (i.e., $V \in [-50,-15]$ km~s$^{-1}$)
as can can be seen in Fig.\ref{diffspectra} and Fig.\ref{opa}.
In contrast, the flat absorption trough at $V \in [-7,+12]$
km~s$^{-1}$ shows only small intensity variations (Fig.\ref{flux}b).
%Those, as we have seen, primarily reflect the variations
%of the NE continuum source intensity relative to the SW one.

The major changes are observed on a time scale of 2 to 3 years. In contrast, the
variations of the absorption trough near 0 km~s$^{-1}$, which typically reach $\simeq \pm
20\%$ of the SW source flux ( see Fig. \ref{flux}), have much shorter time scales: less
than 2 months. As mentioned above, they mostly reflect variations in the NE/SW source flux
ratio, caused by the difference in path lengths between the NE and the SW images
(\citealt{lov98,wik99}), variations which tend to be smeared out after a few months.

Whereas the short term variations of the absorption near 0 km~s$^{-1}$ are caused by sudden
flux variations of the background quasar, coupled to the propagation delay just mentioned,
the reason for the large long term variations are less clear.

We first note that the destruction (or formation) of the HCO$^+$ and HCN molecules on time
scales of months are out of question in view of the large extent of the absorbing cloud(s)
that we derive below ($\sim 1$ pc). For the same reason, significant changes of the HCO$^+$
and HCN line excitation temperature can be ruled out.

To explain the growth of the $-146$ km~s$^{-1}$ absorption feature in 1998-1999 and its
fade-out in 2006-2007, we have to involve either
{\it a)} a motion of the clouds causing the $-146$ km~s$^{-1}$ absorption with respect to
the background continuum source, {\it b)} a micro-lensing event in the spiral galaxy, or
{\it c)} the appearance/disappearance of a bright spot in the quasar. 

The monitoring of the total flux and flux ratio (Fig.\ref{flux}) is instructive in this respect. The
total flux first increased from 1 Jy to 2 Jy between 1997 and mid-1999 , stood well above 2 Jy for a
period of $2-3$ yr, and decreased from 2.5 Jy to 1.5 Jy between 2002 and 2003. These flux changes are
in phase with the absorption changes just described, which seems hardly compatible with a simple
motion of the clouds with respect to the source as in {\it a)}. By 2007, the total flux has recovered
its 2001 value of $2.5-3$ Jy, whereas the $-146$ km~s$^{-1}$ absorption feature has remained
vanishingly weak. We note that \cite{lew03} have tentatively interpreted absorption-line variations
in terms of micro-lensing. However, the good correlation between the NE $-146$ km~s$^{-1}$ and the
SW $-20$ km~s$^{-1}$ blue wing absorption variations, plus the relatively stable value of the flux
ratio over 1996-2007, seem to rule out micro-lensing events as in {\it b)}. The most likely cause
of the absorption variations seems thus to be {\it c)}, the recurrent emission/disappearance of bright
emission blobs in the quasar.

\section{Discussion}

\subsection {The continuum source size}

The structure of PKS~1830$-$211 has been studied with VLBA interferometry at frequencies of
up to 43~GHz (see \citealt {jin03} and references therein). At 43~GHz, the continuum
emission reduces to two compact sources of similar sizes, the NE and SW images, plus two
arc-like features that are the remnants of the Einstein ring observed at longer
wavelengths. The compact sources probably image the optically thick radio core of the
blazar and the arclets, the jet. The arclet emission has a steeper spectral index than the
core emission and its flux at 43~GHz is already 10 times lower than the core flux. It
should thus become negligible at the frequency of our observations (94~GHz). \cite{jin03}
marginally resolve the core images at 43~GHz with a 0.5 mas synthesized beam and find, for
the NE core, a size of $\simeq 230 \times 150$ $\mu$as that corresponds to $1.8 \times 1.2$
pc in the plane of the $z=0.89$ galaxy (assuming a flat Universe with the standard
cosmology parameters: H$_0 = 70$ km~s$^{-1}$~Mpc$^{-1}$, $\Omega_M = 0.3$ and
$\Omega_\Lambda = 0.7$).
%the angular size scale is 7.8 kpc/arcsec at a redshift $z=0.88582$.
According to VLBI data, the core image scales almost linearly in size with wavelength, as
expected for a self-absorbed synchrotron source. Therefore, at 94~GHz, its diameter should
be $0.5-1$ pc.

\subsection {The absorbing clouds in front of the NE source}

The $-146$ km~s$^{-1}$ HCO$^+$ (2-1) absorption reached a depth of $\sim
20\%$ of the total continuum in 1999, when the flux of the NE image was 62\% of the total
flux. This, and the VLBA core size measurements, set strong constrains on the peak optical
depth $\tau_0$, as well as on the source filling factor $f_c$ and absorbing cloud sizes.

We have:
\begin{equation}
\tau_0 = -\ln{\left [ 1- \frac{I_0-I_A}{f_c I_{NE}} \right ]},
\end{equation}
\noindent where $I_0$ is the continuum level ($I_0= 1$ for the normalized spectra),
$(I_0-I_A)$ the apparent depth of the absorption for the $-146$ km~s$^{-1}$ feature ($= 0.2
I_0 $), and $I_{NE}= 0.62 I_0$ the NE image continuum flux. Taking into account that $f_c
\leq 1$ and $\tau \leq \infty$, we find from Eq.[1] that $\tau_0 \geq 0.4$ and $f_c \geq
1/3$. Thus, in 1999, the absorbing material was covering at least one-third of the NE
image.

MGD noted a great similarity in the shapes of the $-146$ km~s$^{-1}$ absorption feature
in HCO$^+$ and HNC, despite the 4 times weaker intensity of the HNC feature. This similarity,
and the non-detection of the $-146$ km~s$^{-1}$ feature in H$^{13}$CO$^+$, down to a level
50 times lower than HCO$^+$, probably means that the clouds responsible for the NE absorption
cannot be optically thick. Adopting this view, we find $\tau_0 \simeq 0.4$ and $f_c \simeq 1$.

How many clouds are actually responsible for the absorption of the NE core and what can we
learn about them? Obviously, the number ought to be small, judging from large amplitude of
the opacity variations, as well as from the shape of the absorption profile.

We have fitted the profiles of the $-146$ km~s$^{-1}$ feature, observed at different epoch, with
Gaussians. Within the noise, all profiles can be fitted with 3 Gaussians of HPW $5-8$ km~s$^{-1}$.
% The 3 gaussians are centred $\pm 1$ km~s$^{-1}$ at $-153$, $-146$ and $-141$ km~s$^{-1}$,
respectively. Following \cite{mar94}, we have also carried out a statistical analysis of the
profile variations assuming a random distribution of absorbing clouds. The clouds
are assumed to have identical column densities, but a size spectrum $dN/dL \propto L^{-(D+1)}$,
where we adopted for the fractal dimension $D = 1.5$. Within the limits of this exercise (limited
period of monitoring, integration over the extent of the continuum sources, and, mostly,
scarcity of information on the motion of these sources, as will be seen below), we find that the
observed large opacity variations imply a small number of absorbing clouds, $N_{abs}$. More
specifically, we find in the framework of Marscher \& Stone's model that the expected fractional
change in equivalent width, $\langle \delta W \rangle / \langle W \rangle$ exceeds 50\% only for
$N_{abs} \leq 5$ \footnote{This number was derived from equation (7) of \cite{mar94}, after
correction of a typographical error on the last exponent, which should read $n+i-1$ instead of $n$}.
Then, the clouds responsible for the bulk of the absorption should not be much smaller than the
size of the background continuum source, $\sim 1$ pc. Of course, this simple analysis is biased
against structures much smaller than the continuum sources and we cannot rule out the presence of
small cloudlets whose cumulative contribution would be a small fraction of the absorption.

From the observation of two different rotational transitions, \cite{wik96} and \cite{car98}
derived rotation temperatures $T_{rot}$ for CS, H$^{13}$CO$^+$, H$_2$CO and N$_2$H$^+$ (and
upper limits to those of H$^{12}$CO$^+$ and HCN) close to (slightly higher than) the
expected cosmic background radiation temperature, T$_{CBR} = 5.1$ K at $z=0.89$, implying
that the average gas density is $<10^5$ cm$^{-3}$. Unfortunately, the first 3 rotational
transitions of CO are unobservable from the ground, and only the J= 4-3 line has been so
far detected (\citealt{wik98,ger97}), so we do not know the rotation temperature of this
low dipole moment species and cannot set a more constraining upper limit to the gas
density. In any case, all multi-transition observations have been made toward the SW
source, where absorption is stronger, and there is no direct constraint on the gas density
toward the NE source.

The CBR temperature, however, is high enough to populate the HCO$^+$ $J=1$ level ($E/k = 4.3$ K)
without the help of collisions. At radiative equilibrium, the HCO$^+$ $J=2-1$ line opacity
becomes $\simeq 0.4$ as soon as the HCO$^+$ column density per km~s$^{-1}$ reaches $dN/dv =
10^{12}$ cm$^{-2}$~km~s$^{-1}$, i.e. the H column density $N({\rm H})$
exceeds a few $\times 10^{21}$ cm$^{-2}$, assuming the fractional HCO$^+$ abundance is similar to that in
diffuse Galactic clouds ($2 \times 10^{-9}$, according to \citealt{lis00}). For a cloud depth of 1 pc, the gas
density in the NE absorbing clouds would be $n({\rm H})= {\rm few}
\times 10^2$ cm$^{-3}$ and the visual extinction per cloud, presumably, $A_v \sim 1$
mag, if the dust properties are similar to those in the vicinity of
the Sun: the absorbing cloud(s) look very much alike the local
Galactic diffuse clouds.

In the case of PKS~1413+135, \cite{wik97} find excitation temperatures for the $J=1-0$, $2-1$
and $3-2$ lines significantly higher than the 3.4 K CBR temperature at $z=0.25$, and which
seem to increase with increasing $J$. They interpret their data as evidence of at least two
gas components along the line of sight, one of which could be in the form of dense clumps.
However, the line of sight to PKS~1413+135 crosses a much larger fraction of the molecular
disk than in the case of PKS~1830$-$211, since the galaxy is seen almost edge-on, and the
solid angle subtended by the continuum source is probably smaller, since PKS~1413+135 is
not magnified, two conditions that may enhance the visibility of small clumps.

\subsection{Variability of the continuum source}

The variations in the absorption profiles cannot be caused by the proper motion of the $z=0.89$
galaxy with respect to the observer. Indeed, the apparent drift of the clouds with respect to
the continuum source is:
\begin{equation}
s = V_\perp \Delta {\rm t} / (1+z_{abs}),
\end{equation}
\noindent where $V_\perp$ is the transverse velocity and $\Delta$t the time interval in the Earth
frame, for which we apply the relativistic correction for the time dilation measure at the redshift
$z_{abs}$ of the galaxy. With $V_\perp \sim 1000$ km~s$^{-1}$ and on time scales of about a year,
$s \sim 100$ AU, i.e., 3 orders of magnitude smaller than the extent of the clouds derived above
($\geq 0.5$ pc).

Whereas the observer, the absorbing clouds and the background galaxy that host the quasar
should have relative velocities $\la 1000$ km~s$^{-1}$, the images of the radio core,
which consists of plasmons, can move arbitrarily fast, provided the geometry is favorable.
VLBA measurements at 43~GHz show variations in the apparent separation between the NE and SW
cores on time scales of only months. Early in 1997, \cite{jin03} measured this parameter 8 times
at intervals of 14 days or so. These authors found that the NE core source moved with respect to
its SW counterpart by as much as $\simeq 140$ $\mu$as in 4 months. A comparison with previous data,
obtained 8 months earlier, showed a change $\sim 200$ $\mu$as in the opposite direction. This is
about the extent of the continuum cores at 43~GHz and twice their expected extent at 94~GHz.
Projected on the plane of the lensing galaxy, 200 $\mu$as corresponds to $\sim 1.6$ pc. We note
that the relative displacement of the NE and SW sources implies an apparent velocity of $v \sim 8c$
in this plane, so that the source/observer geometry should indeed be particularly favorable.
There are more high frequency data on PKS~1830$-$211 in the VLBA archive. Once calibrated,
they may teach us about the motion of the continuum sources throughout our monitoring period.
This will be the object of a subsequent study.

\cite{nai05} interpret the apparent motion of the NE and SW sources by the recurrent
ejection of plasmons along a helical jet. The plasmons are ejected at relativistic
velocities by a precessing "nozzle" pointing toward the observer. The plasmons appear as
bright self-absorbed synchrotron sources and remain visible at millimeter wavelengths for a
few months. Traces of fading plasmons may have been observed by \cite{gar97} in July 1996
at 43~GHz, in the form of compact emission "knots" located close to the bright NE and SW
cores. These knots had vanished by January 1997, when \cite{jin03} resumed VLBA
observations. Unfortunately, there is no higher frequency data allowing us to zoom deeper
into the barely resolved cores. 

We have noted above the good correlation between the opacity variations of the $-146$ km~s$^{-1}$
NE and $-30$ km~s$^{-1}$ SW absorption components on Fig.\ref{opa}. Obviously, these components
are two clouds observed against the NE and SW images of the same plasmon. That both components
increased and faded out at the same time (within the coarse sampling of our observations) may mean
that the bright plasmon responsible of the bulk of the absorption at these velocities appeared in
1999 and faded out in 2003. The total flux variations (Fig.\ref{flux}a) seem to corroborate
this picture. The new increase of the continuum flux in 2007, which seems to have not affected the
absorption profiles, could be due to the appearance of a new plasmon in a direction free of any cloud.

\subsection{The nature of the optically thick absorption component toward the SW source}

As pointed out by several authors, the wide range of velocities ($\simeq 100$ km~s$^{-1}$) over which
absorption is seen toward the SW core image is puzzling: the spiral galaxy appears almost face-on in
the HST optical images, and the SW core source is very small ($\sim 1$ pc). How can such a large span
of velocities be reached over such a small region? According to gravitational lens models of
PKS~1830$-$211 (e.g. \citealt{nai93}), the SW core source image lies $2-3$ kpc away from the galactic
nucleus, so that no large velocity gradient is expected over an area of a few pc. Could the absorptions
near/below $-30$ km~s$^{-1}$ and above +30 km~s$^{-1}$ come from a distinct, more extended continuum
source filtered out in the VLBI observations? Probably not, as there is no trace of such a source on
the VLA 47~GHz maps of \cite{car98}. 
Moreover, the HCO$^+$ $J=2-1$ observations of MGD in the PdBI extended configuration show
that the absorption in the SW component wings arises within $\sim 1$ kpc from the SW core source.

The width of the absorption and the detection toward the SW source of very rare isotopologues such as
HC$^{17}$O$^+$ and HC$^{15}$N imply molecular column densities and masses characteristic of large
Galactic molecular clouds: H$_2$ column density of $3 \times 10^{23}$ cm$^{-2}$ and molecular gas mass
integrated over 1 square parsec of $\simeq 5 \times 10^3$ M$_{\odot}$, if we assume fractional abundances
relative to H$_2$ of $10^{-9}$ for HCN and HCO$^+$ and $10^{-7}$ for OH (see MGD and \citealt{lis99}).
We note that the H$_2$ column density we adopt is higher than those derived by \cite{wik98} and \cite{ger97}
from CO(4-3) observations. However, the high energy of the CO J=4 level ($E/k=56$ K) and the rather low
signal-to-noise ratio of the CO spectra make the latter column densities fairly uncertain.
%extent of cloud large because absorption has not disappeared?
Even so, the very broad absorption is more reminescent of the Galactic Center clouds (e.g., \citealt{chr05})
or of exceptional Giant Molecular Clouds such as W49A-N (\citealt{wil04}) than of the more common molecular
clouds scattered in the arms of the Galactic disk. Is the model of the lensing galaxy inaccurate and could
the SW source be closer to the center of the $z=0.89$ galaxy? Higher angular resolution observations of the
HCO$^+$ $J=2-1$ absorption would be warranted to solve this question.

We address a final point concerning the implication of the time variations for our previously published
study of isotopic abundance ratios in the $z=0.89$ galaxy (MGD). The sole reason for us to monitor the
HCO$^+$ (2-1) absorption profile was to check for changes in the NE/SW flux ratio that could have biased
the interpretation of our measurements. Fortunately, no significant changes seem to have occured between
1999 and 2001, when most of our data on rare isotopes were taken.
%{\em J'hesite a mettre 2002, car l'aile de la raie vers -20 km~s$^{-1}$ a deja change ...
%ca pourrait de fait changer un peu notre 12/13C ainsi que les rapports associes 14/15N et 16/18O. *** OK***}
The H$_2^{32}$S/H$_2^{34}$S ratio ($= 10.6 \pm 0.9$) was measured later in 2005-2006, but the lines of
both isotopologues are close in frequency and were observed simultaneously. The new PdBI receivers, with
larger bandwidth and flexible correlator setting, now allow us to simultaneously observe lines distant in
frequency by as much as 4~GHz. This is the case for example for the $J=2-1$ lines of H$^{13}$CO$^+$,
HC$^{18}$O$^+$ and HC$^{17}$O$^+$ in sources like PKS~1830$-$211 or B~0218+357 (Muller et al., in prep.).

\section{Conclusions}

Our monitoring of the absorption at $z=0.89$ on the line of sight to the quasar
PKS~1830$-$211 has revealed spectacular changes in the HCO$^+$ and HCN (2-1) line profiles
on time scales of about a year. These primarily concern the absorption toward the
quasar NE image, which increased by a factor of $\simeq 3$ between 1996 and 1999 and
decreased by a factor $\geq 6$ between 2003 and 2006. They also concern the broad
absorption wings observed toward the SW image.

We interpret the large absorption variations as resulting from the appearance/disappearance
of discrete components in the background continuum source. Indeed, the good correlation
between the opacity variations toward the NE and the SW images seem to rule out
micro-lensing events. The continuum source most likely consists of bright plasmons
sporadically ejected at relativistic velocities, a motion magnified by the gravitational
lens. The separation between the NE and SW continuum images has been found to vary between
1996 and 1997 by $\sim 0.2$ mas from VLBA observations. Assuming that a motion of similar
amplitude occurred in 2003, the size of the continuum images, the depth of the absorption,
and the near disappearance of the NE absorption feature since 2006 teach us that the
absorbing clouds toward the NE source should be: {\it a)} sparse, as otherwise the
variations would be less pronounced, {\it b)} compact, as they must have covered at some
point most of the 1 pc continuum image and {\it c)} have sizes covering a good fraction of
a pc. It is likely that these clouds are diffuse clouds. The clouds which are origin of the
saturated absorption toward the SW source, on the other hand, are much more massive and
should be part of a Giant Molecular Cloud complex.

So far, there have been no VLBI observations of the HCO$^+$ and/or HCN line absorptions.
Such observations at a resolution $\leq 0.1$ mas would help to clarify the morphology of
the clouds and may explain why the absorption toward the SW image is so extended in
velocity. If the strong absorption changes observed in 1999 and 2003 are recurrent, coupled
millimeter-wave absorption measurements and mm-VLBI continuum observations may help to
better constrain the density and sizes of the molecular clouds in the $z=0.89$ galaxy,
stressing the potential of molecular absorption for the exploration of the interstellar
medium in distant objects.

\begin{acknowledgements}
We would like to thank F. Combes and T. Wiklind for kindly communicating us their 30-m
telescope data. We also thank F. Combes, H. Liszt and G. Henri for helpful discussions
and the PdB Observatory staff and IRAM-Grenoble SOG for their support in the observations.
We acknowledge the referee for giving helpful and constructive comments.
MG would like to thank the Taiwanese National Science Council for its support during a stay in ASIAA.
Based on observations carried out with the IRAM Plateau de Bure Interferometer.
IRAM is supported by INSU/CNRS (France), MPG (Germany) and IGN (Spain).
\end{acknowledgements}

\end{document}